# Day-Ahead Solar Forecasting Based on Multi-level Solar Measurements

Mohana Alanazi, Mohsen Mahoor, Amin Khodaei
Department of Electrical and Computer Engineering
University of Denver
Denver, USA
mohana.alanazi@du.edu, mohsen.mahoor@du.edu, amin.khodaei@du.edu

*Abstract*—The growing proliferation in solar deployment, especially at distribution level, has made the case for power system operators to develop more accurate solar forecasting models. This paper proposes a solar photovoltaic (PV) generation forecasting model based on multi-level solar measurements and utilizing a nonlinear autoregressive with exogenous input (NARX) model to improve the training and achieve better forecasts. The proposed model consists of four stages of data preparation, establishment of fitting model, model training, and forecasting. The model is tested under different weather conditions. Numerical simulations exhibit the acceptable performance of the model when compared to forecasting results obtained from two-level and single-level studies.

*Keywords*—Solar generation forecast, nonlinear autoregressive with exogenous input (NARX).

## NOMENCLATURE

| | |
|---|---|
| $P_{actual}$ | Actual solar generation |
| $P_{forecast}$ | Forecasted solar generation |
| $\bar{P}_{actual}$ | Average actual solar generation |
| $N$ | Number of sample |
| $E_c$ | Calculated error at customer level |
| $E_f$ | Calculated error at feeder level |
| $E_S$ | Calculated error at substation level |

## I. INTODUCTION

SOLAR FORECASTING plays a key role in the planning, control and operation of power systems. Although this viable generation technology is making fast inroads in electricity grids, solar forecasting is still facing various challenges, due to the inherent variability and uncertainty in solar photovoltaic (PV) generation [1], [2].

Numerous factors, including but not limited to the dropping cost of solar technology, environmental concerns, and the state and governmental incentives, have made the path for a rapid growth of solar regeneration. More than 2 GW of solar PV was installed only in the U.S. in the summer of 2016, which is 43% higher compared to the installed capacity in the same timeframe in 2015, to achieve an accumulated capacity of 31.6 GW [3]. Accordingly, solar forecasting problem has attracted more attention to properly incorporate solar generation into power system planning, operation, and control.

Many research studies are carried out on solar forecasting problem, and several approaches are suggested to improve forecasting results [4]-[8]. In [9], a short-term one-hour-ahead Global Horizontal Irradiance (GHI) forecasting framework is developed using machine learning and pattern recognition models. This model reduces normalized mean absolute error and the normalized root mean square error compared to the commonly-used persistence method by 16% and 25%, respectively. In [10], an intelligent approach for wind and solar forecasting is proposed based on linear predictive coding and digital image processing. It is shown that the model can outperform conventional methods and neural networks. Ensemble methods are quite popular in statistics and machine learning, as they reap the benefit of multiple predictors to achieve not only an aggregated, but also a better and reliable decision. A survey paper on using ensemble methods for wind power forecasting and solar irradiance forecasting is proposed in [11]. The paper concludes that the ensemble forecasting methods in general outperform the non-ensemble ones. A comprehensive review focusing on the state-of-the-art methods applied to solar forecasting is conducted in [12]. A variety of topics including the advantages of probabilistic forecast methods over deterministic ones, and the current computational approaches for renewable forecasting is discussed in this paper.

Historical data are of great importance for solar forecasting. By leveraging historical data, the solar PV generation can be forecasted for various time horizons as discussed in [13]. This study investigates least-square support vector machines, artificial neural network (ANN), and hybrid statistical models based on least square support vector machines with wavelet decomposition. In addition, a variety of measures, including the root mean square error, mean bias error and mean absolute error, are employed to evaluate the performance of the aforementioned methods. The hybrid method based on least-square support vector machines and wavelet decomposition surpasses the other methods. A new two-stage approach for online forecasting of solar PV generation is proposed in [14]. This approach leverages a clear sky model to achieve a statistical normalization. Normalized power output is further forecasted by using two adaptive linear time series models; autoregressive and autoregressive with exogenous input. Various types of ANN, including but not limited to recurrent neural network, feed-forward neural network, and radial basis function neural network are employed for solar forecasting.

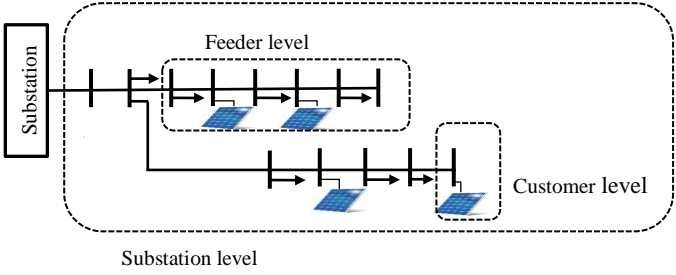

Fig. 1 Multi-level Solar PVs installed at different locations.

The ANNs not only can process complex and nonlinear time series forecast problems, but also can learn and figure out the relationship between the input and the target output. On the basis of ANN, a statistical method for solar PV generation forecasting is proposed in [15]. One of the lessons learned from this paper is that neural networks can be well-trained to enhance forecast accuracy. In [16], by levering stationary data and employing post-processing steps, a feed-forward neural network-based method for day ahead solar forecasting is studied. A comprehensive review of solar forecasting by using different ANNs is provided in [17].

Hybrid models are considered highly effective for solar forecasting in a way that they reinforce capabilities of each individual method. Hybrid models reap the benefits of two or more forecasting methods with the objective of achieving a better forecast result [18]-[21]. In [22], authors present a hybrid model consisting of various forecasting methods for a 48-hour-ahead solar forecasting in North Portugal. This study advocates that the hybrid model attains a significant improvement compared to statistical models. Another hybrid short-term model to forecast solar PV generation is studied in [23]. This hybrid model is formed on the basis of both group method of data handling and least-square support vector machine, where the performance of the hybrid model significantly outperforms the other two methods.

The existing literature in this research area lacks studies on multi-level data measurements for day-ahead solar PV generation forecasting. Leveraging the multi-level solar measurements to provide a more accurate forecasting for the solar PV generation is the primary objective of this paper. The solar PV generation, which is measured at various locations including customer, feeder and substation, is utilized for day-ahead solar forecasting with the objective of enhancing the forecast accuracy. These multi-level measurements could play an instrumental role in enhancing solar forecasting in terms of reaching lower error values. The proposed forecasting model, which will be further discussed in details in this paper, consists of four stages and takes advantage of multiple datasets related to specific locations.

The rest of the paper is organized as follows: Section II discusses outline and the architecture of the proposed forecasting model. Numerical simulations are presented in Section III. Discussions and conclusions drawn from the studies are provided in Section IV.

## II. FORECASTING MODEL OUTLINE AND ARCHITECTURE

Fig. 1 depicts the three levels of solar PV measurements: customer, feeder, and substation. The proposed model aims to outperform the forecast applied at each solar measurement level. The forecast in each level is performed using a nonlinear autoregressive neural network. The mean absolute percent error MAPE is accordingly calculated as in (3) for each level, and denoted as $E_C$, $E_F$, and $E_S$ for customer, feeder, and substation, respectively. This model aims to reduce the forecasting error to be less than the minimum of $E_C$, $E_F$, and $E_S$. Fig. 2 depicts the three datasets, which are processed under different stages and explained in the following.

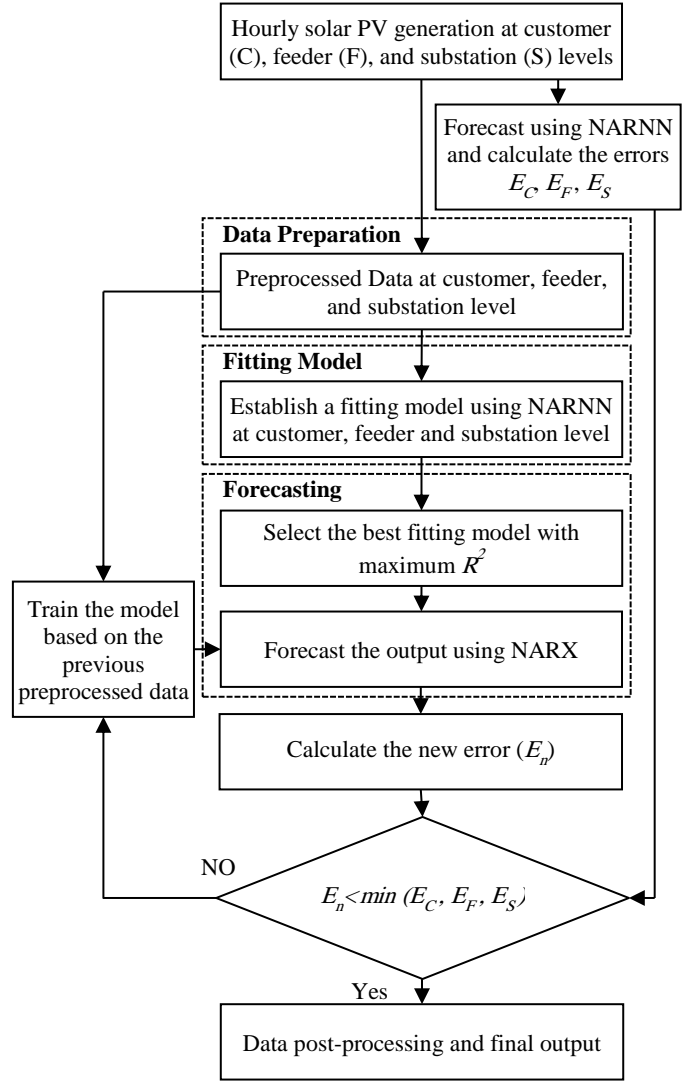

Fig. 2 The flowchart of the multi-level solar PV generation forecasting.

### A. Data Preprocessing and Adjusment

The data used in the simulation represent the total solar PV generation. The data preparation includes removing offset, normalization, removing nighttime values, and stationarization. More detail about data preprocessing can be found in [16]. The data preparation is to ensure the quality of dataset before it is inputted to the forecasting model. This step includes the simulation of maximum power generated from solar PV at clear sky conditions. This is achieved by simulating the maximum solar PV generation at clear sky conditions using the system advisory model (SAM) provided by National Renewable Energy Laboratory (NREL) [24]. The maximum solar irradiance along with different metrological inputs in clear condition are fed to SAM in order to simulate the maximum solar PV generation. Fig. 3 presents the flowchart for data preparation.

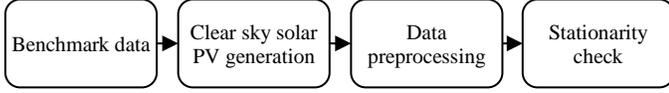

Fig. 3 The flowchart for data preparation.

*B. Fitting Model*

By using NARNN, the fitting model is created for each level. In this respect, the NARNN model utilizes a large set of historical data in order to train the model and then forecast the output. It is applied to the three datasets, including customer, feeder, and substation in order to establish the three fitting models. The best fitting model among the three is selected using the coefficient of determination $R^2$. The coefficient of determination examines the proportion variance of the predicted fitting model. The coefficient of determination can be expressed mathematically as in (1), where $\overline{P(t)}_{actual}$ is the average of the actual data over the number of sample. The $R^2$ ranges from 0 to 1, where 0 represents that the fitting model is not predictable, and 1 means that the NARNN is able to predict the fitting without any error. So, the best selected fitting model among the three is the one with maximum $R^2$.

$$R^2 = 1 - \left[ \frac{\sum_{t=1}^{N}(P(t)_{actual} - P(t)_{forecast})^2}{\sum_{t=1}^{N}(P(t)_{actual} - \overline{P(t)}_{actual})^2} \right] \quad (1)$$

*C. Forecasting*

NARX is a time series model that predicts the output using historical values $y(t)$ as well as inputs $x(t)$. The NARX model is presented in (2), where $d$ is the number of considered historical values. The fitting model is fed as an input to NARX along with the previously preprocessed data. The NARX is trained and the output is forecasted. Fig. 4 depicts the architecture of the NARX. The goal is to forecast a day-ahead solar PV generation with a new error $E_n$, which is less than the minimum of the three errors as shown in the flowchart in terms of a condition.

$$y(t) = f(x(t-1),..,x(t-d), y(t-1),..,y(t-d)) \quad (2)$$

*D. Data Post-processing*

The output from the forecasting model is post-processed by denormalizing, adding nighttime values, and calculating the final solar output as explained in detail in [16]. MAPE and the root mean square error (RMSE) are calculated as in (3) and (4), respectively.

$$MAPE = \frac{1}{N} \sum_{t=1}^{N} \left| \frac{P(t)_{actual} - P(t)_{forecast}}{P(t)_{actual}} \right| \quad (3)$$

$$RMSE = \sqrt{\frac{1}{N} * \sum_{t=1}^{N}(P(t)_{actual} - P(t)_{forecast})^2} \quad (4)$$

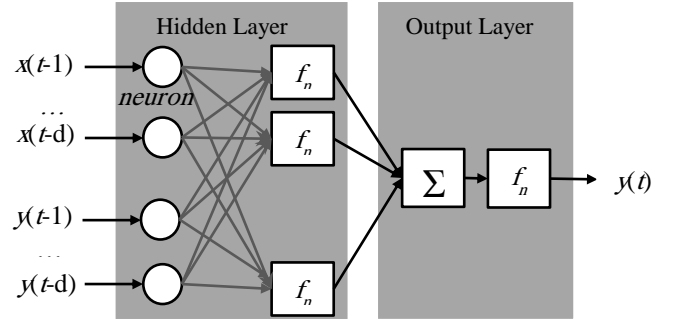

Fig. 4 The architecture of the NARX

### III. NUMERICAL SIMULATIONS

The hourly solar PV generation of three levels including customer (C), feeder (F), and substation (S), for a specific area in Denver, Colorado are utilized to perform forecasting. The data used in this model are available in [25]. The customer level data are considered as the aggregated customers' solar PVs generation for a selected area. The feeder level data are the aggregated solar PVs generation for each feeder, in which four feeders are considered in this study. Finally, the substation level data are the solar PV generation measured at the substation level. In order to demonstrate the merits of the proposed model, the following three cases with various weather conditions are investigated:

**Case 1:** Forecast using NARNN for each level without data processing.

**Case 2:** Forecast using NARX with three-level measurements and data processing.

**Case 3:** Forecast using NARX with two-level measurements and data processing.

**Case 4:** Forecast using NARX with single-level measurement and data processing.

**Case 1:** In this case, by leveraging NARNN, day-ahead solar PV generation is forecasted for all three levels, while ignoring data processing. The calculated MAPE and RMSE for the customer, feeder, and substation levels with different weather conditions are listed in Table I. As highlighted in Table I, the customer level forecast has achieved the minimum MAPE as well as RMSE for the selected weather conditions. This case is considered as a base case, in which the calculated values are utilized in order to demonstrate the effectiveness of using the three-level measurements for forecasting. The objective in the following case is to apply the proposed model in order to get a new error that is less than the minimum achieved under this case.

TABLE I
CASE 1: MAPE AND RMSE FOR THE CONSIDERED DATASETS UNDER DIFFERENT WEATHER CONDITIONS

| Dataset level | Sunny | | Cloudy | | Partly Cloudy | |
|---|---|---|---|---|---|---|
| | RMSE (kW) | MAPE (%) | RMSE (kW) | MAPE (%) | RMSE (kW) | MAPE (%) |
| Customer | 44.58 | 4.47 | 20.54 | 6.04 | 36.11 | 4.09 |
| Feeder | 48.81 | 8.29 | 23.12 | 7.34 | 38.59 | 6.73 |
| Substation | 70.77 | 10.41 | 43.65 | 10.13 | 53.03 | 10.37 |

**Case 2:** In this case, three-level measurements are preprocessed in order to ensure the quality of the training data fed to the forecasting model. This case includes three forecasting stages: establishing the fitting model from each measurement level using the NARNN, training the NARX model using the previously preprocessed datasets, and forecasting the solar PV generation using the three-level measurements and the fitting model as input. The fitting model with the minimum MAPE and the maximum $R^2$ is selected as input to NARX. Table II exhibits how well the fitting model is established in terms of $R^2$ and MAPE for the three-level measurements under different weather conditions. As highlighted in Table II, the fitting model established by using the customer level measurement outperforms the ones established by using the feeder and substation measurements. In order to show the merit of using three-level measurements for the same location, the three measurements along with the best fitting model are fed as inputs to NARX to forecast the solar PV generation. The forecast is simulated for the same selected days in Case 1. Table III exhibits the MAPE and RMSE for the selected days. The forecast errors in this case are less than the minimum achieved in Case 1. Fig. 5, 6 and 7 depict the forecasted and actual solar PV generation for the considered sunny, cloudy, and partly cloudy days, respectively.

TABLE II
THE FITTING MODEL MAPE AND $R^2$ FOR THE CONSIDERED LEVELS UNDER DIFFERENT WEATHER CONDITIONS

| Dataset level | Sunny | | Cloudy | | Partly Cloudy | |
|---|---|---|---|---|---|---|
| | MAPE (%) | $R^2$ | MAPE (%) | $R^2$ | MAPE (%) | $R^2$ |
| Customer | 2.39 | 0.9987 | 2.29 | 0.9956 | 3.49 | 0.99 |
| Feeder | 3.78 | 0.996 | 2.80 | 0.9943 | 4.02 | 0.988 |
| Substation | 5.95 | 0.9873 | 4.98 | 0.9821 | 5.37 | 0.986 |

**Case 3**: In this case only two measurements at customer and feeder levels are used for forecasting. The preprocessed data along with the best selected fitting model are fed to NARX. The forecasting performance of this case is shown in Table III.. In sunny day, this case has reduced the MAPE compared to Case 1 by 47%. In cloudy and partly cloudy weather conditions, Case 3 has reduced the MAPE compared to Case 1 by 61% and 19%, respectively.

**Case 4:** To exhibit the effectiveness of the three-level measurements, Case 2 is repeated, but only one measurement (customer level) is included as an input to NARX. Similar to the previous case, the best fitting model based on MAPE and $R^2$ is fed to NARX along with preprocessed customer level measurement. Table III shows the forecast error using NARX with single-level measurement comparing to NARX with three-level measurements, two-level measurements, and the minimum forecast error among the single-level measurement using NARNN without data processing. A single-level measurements considerably improve the results over NARNN method, however achieve not as good solution as in two previous cases with three- and two-level measurements.

TABLE III
THE MAPE FOR DIFFERENT CASE STUDIES

| Weather Condition | Minimum MAPE (Using NARNN without data processing) | MAPE (Using NARX and three-level processed data) | MAPE (Using NARX and two-level processed data) | MAPE (Using NARX and single-level processed data) |
|---|---|---|---|---|
| Sunny | 4.47 | 1.67 | 2.38 | 3.14 |
| Cloudy | 6.04 | 2.10 | 2.36 | 2.44 |
| Partly Cloudy | 4.09 | 2.69 | 3.30 | 3.39 |

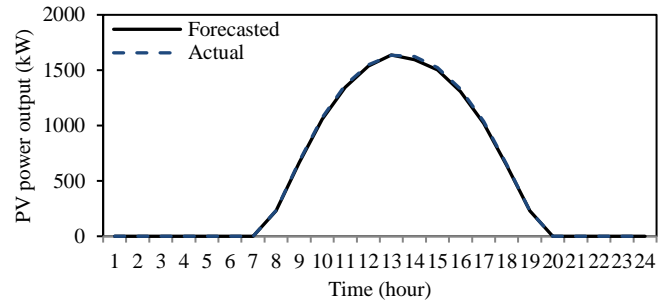
Fig. 5 Actual and forecasted solar PV generation in a sunny day

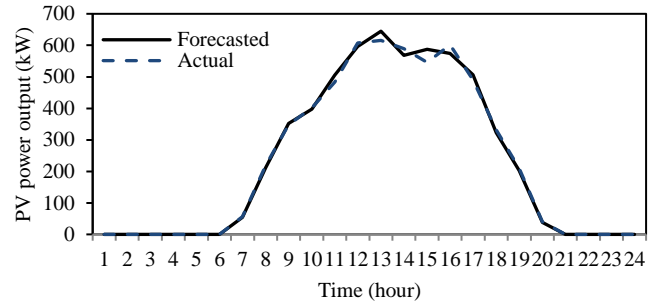
Fig. 6 Actual and forecasted solar PV generation in a cloudy day

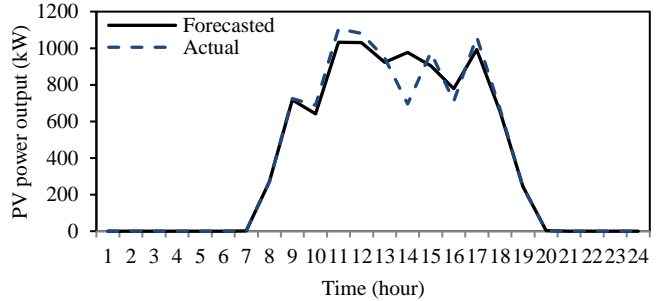
Fig. 7 Actual and forecasted solar PV generation in a partly cloudy day

As shown in Table III, the model minimizes the forecast error to outperform the minimum error reported at customer level. The proposed model has reduced the error compared to the minimum error in Case 1 by 63%, 65%, and 34% for sunny, cloudy, and partly cloudy weather conditions, respectively. Moreover, the merit of using three-level measurements is shown by comparing the forecast error using the proposed model with applying two-level measurements to the model as in Case 3. The MAPE is reduced by 30%, 11%, and 18% for sunny, cloudy, and partly cloudy weather conditions, respectively. The three-level measurement also outperforms Case 4 in which only single-level measurement are included. The three-level measurements model has reduced the MAPE by 47%, 14%, and 21% for sunny, cloudy, and partly cloudy weather conditions, respectively. The previous cases have shown that forecasting performance is greatly impacted by the historical data used to train the model. Multiple historical data for a specific location along with an appropriate data processing will improve the training step and minimize the forecasting error.

## IV. CONCLUSION

In this paper, a day-ahead solar PV generation forecast model based on multi-level measurements was proposed. The proposed model demonstrated an improvement in forecasting accuracy by reducing the MAPE from 14% to 47% for various weather conditions, compared to the case when only single-level measurements were included. It was further seen that the data preprocessing was an important step to ensure the quality of the data before it was used in the training process. The numerical studies revealed that training the forecasting model without data preprocessing might adversely impact the forecasting accuracy. The proposed preprocessing model could potentially reduce the MAPE by 34% to 65%. It was further shown that the three-level measurements help achieve a better forecasting accuracy compared to two-level measurements. The proposed model can be further enhanced by including multiple meteorological parameters such as cloud cover, solar irradiance, and temperature along with three-level measurements as inputs to NARX.